\documentclass[aps,pra,twocolumn,superscriptaddress]{revtex4-1}

\usepackage{graphicx}
\usepackage{dcolumn}
\usepackage{bm}
\usepackage{float}
\usepackage{amsmath}
\usepackage{color}
\usepackage{comment}

\usepackage[pdftex,colorlinks=true]{hyperref}
\hypersetup{
  colorlinks=true,
  linkcolor=blue,
  urlcolor=blue,
}

\usepackage{tikz}
\newcommand*\emptycirc[1][0.4ex]{\tikz\draw (0,0) circle (#1);} \newcommand*\fullcirc[1][0.4ex]{\tikz\fill (0,0) circle (#1);}

\begin{document}
\title{Transversal effects on the ground-state of hard-core dipolar bosons \\ in one-dimensional optical lattices}
\author{Henning Korbmacher}
\affiliation{Institut f\"ur Theoretische Physik, Leibniz Universit\"at Hannover, Appelstr. 2, D-30167 Hanover, Germany}
\author{Gustavo A. Domínguez-Castro}
\affiliation{Institut f\"ur Theoretische Physik, Leibniz Universit\"at Hannover, Appelstr. 2, D-30167 Hanover, Germany}
\author{Wei-Han Li}
\affiliation{Max Planck Institute for the Physics of Complex Systems, Nöthnitzer Str. 38, 01187 Dresden, Germany}
\author{Jakub Zakrzewski}
\affiliation{Instytut Fizyki Teoretycznej, 
Uniwersytet Jagiello\'nski,  \L{}ojasiewicza 11, PL-30-348 Krak\'ow, Poland}
\affiliation{Mark Kac Complex Systems Research Center, Uniwersytet Jagiello{\'n}ski, PL-30-348 Krak{\'o}w, Poland}
\author{Luis Santos}
\affiliation{Institut f\"ur Theoretische Physik, Leibniz Universit\"at Hannover, Appelstr. 2, D-30167 Hanover, Germany}

\date{\today}

\begin{abstract}
Polar lattice gases are usually assumed to have an inter-site interaction that decays with the inter-particle  distance $r$ as $1/r^3$. However, a loose-enough transversal confinement may strongly modify the dipolar decay in one-dimensional lattices. We show that this modification alters significantly the ground-state properties of hard-core dipolar bosons. For repulsive inter-site interactions, the corrected decay alters the conditions for devil's staircase insulators, affecting significantly the particle distribution in the presence of an overall harmonic confinement. For attractive interactions, it results in a reduction of the critical dipole interaction for the formation of self-bound clusters, and for a marked enhancement of the region of liquefied lattice droplets. 
\end{abstract}


\maketitle

\section{Introduction}

Ultracold quantum gases in optical lattices constitute an optimal platform for studying many-body physics under precisely controlled conditions~ \cite{Bloch2008, doi:10.1126/science.aal3837, Langen2015}. In most current experiments, the interactions between particles are short-range and well-modeled by a contact pseudopotential.  However, seminal experiments on dipolar systems formed by magnetic atoms~\cite{dePaz_PRL_111_2013, baier2016extended, Patscheider_PRR_2_2020} and polar molecules \cite{yan2013observation,Li2023} in optical lattices are starting to explore exciting physics beyond the short-range scenario. 
Due to the anisotropic and long-range character of the dipole-dipole potential, polar gases confined in optical lattices are  characterized not only by on-site interactions, but, crucially, also by anisotropic inter-site interactions. As a result, dipolar lattice gases of pinned particles 
can be employed as quantum simulators for spin models, whereas itinerant particles realize different forms of the extended-Hubbard model~\cite{Lahaye2009,Baranov2012}. Compared to their non-dipolar counterparts, dipolar lattice gases present a much richer ground-state physics, including crystalline phases~\cite{Burnell_2009} and  supersolids~\cite{PhysRevLett.104.125301}, or the Haldane-insulator phase~\cite{PhysRevLett.97.260401}.

The spatial decay of the inter-site interactions plays a crucial role in polar lattice gases. Due to the form of the dipolar interaction in free space, this decay is typically assumed as $1/r^3$, with $r$ the inter-site distance. 
However, the interaction decay may be significantly affected by the confinement transversal to the lattice axis~\cite{Wall_2013}. This confinement alters the on-site wave functions, introducing a modification of the inter-site interaction, which may potentially depart very significantly from the $1/r^{3}$ dependence, and hence alter the equilibrium and out-of-equilibrium physics of the polar lattice gas  \cite{Wall_2013, Korbmacher2023}.

In this paper, we show that the modification of the interaction decay in the presence of a loose-enough transversal confinement results in a significant modification of the ground-state properties of hard-core dipolar bosons. In contrast to Ref. \cite{Wall_2013}, which focused on strong transversal confinement, we show that the ground state is much more altered in the experimentally relevant regime of a weak transversal confinement. 
We first consider the case of repulsive interactions, showing that the modified decay results in markedly shifted insulating phases, which translate into a 
distorted particle distribution in the presence of an overall harmonic confinement. For attractive dipoles, we show that the modified decay may significantly ease the conditions for the realization of self-bound clusters. Moreover, it results in a much wider parameter region for the observation of liquefied self-bound droplets~\cite{Morera2023} without the need of super-exchange processes.



\begin{figure}[t]
\centering
\hspace*{0.2cm}\includegraphics[width=0.8\columnwidth]{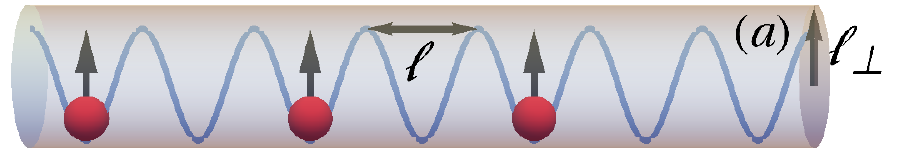}
\hspace*{-0.8cm}\includegraphics[width=0.8\columnwidth]{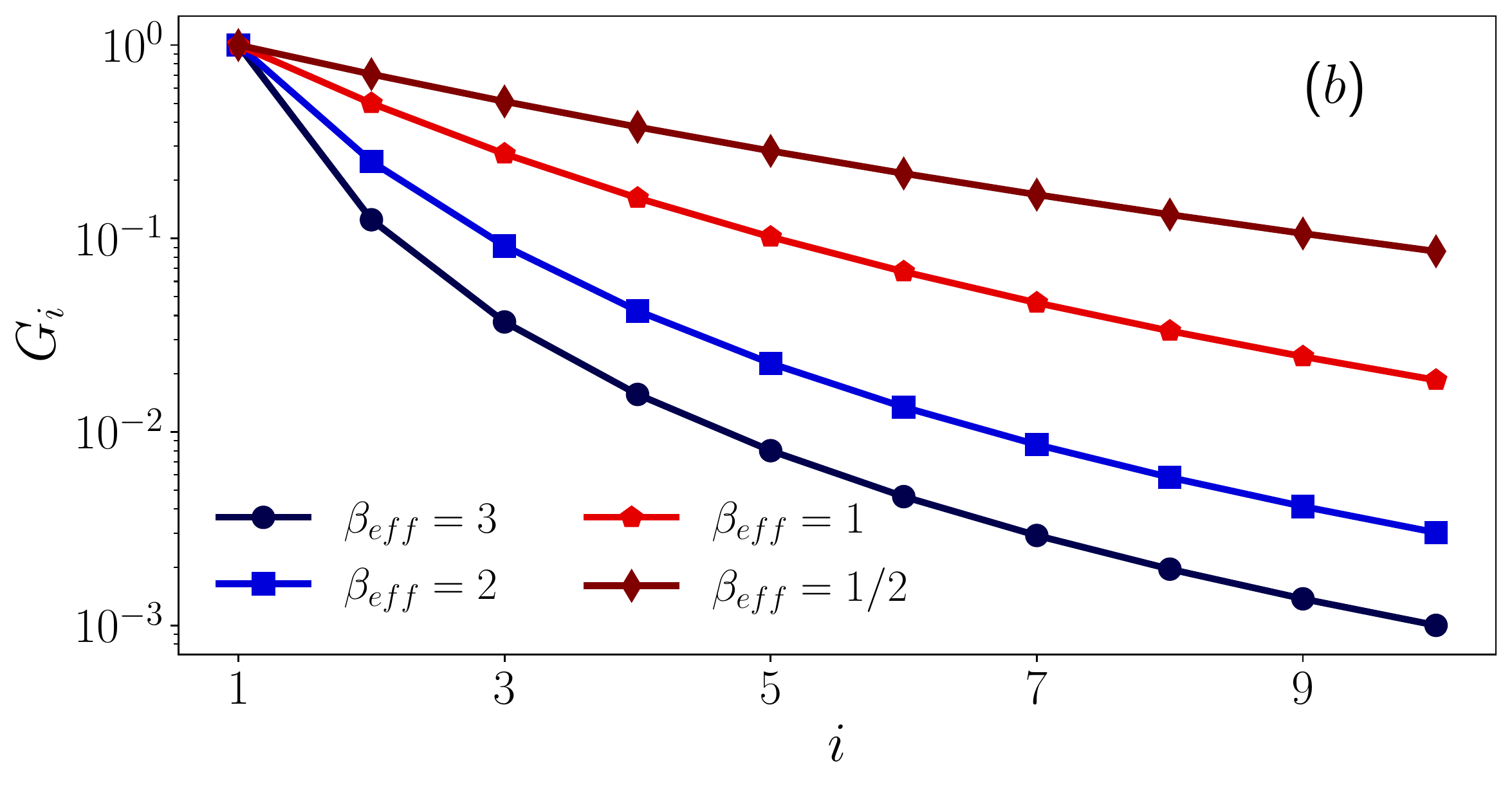}
\caption{(a) Schematic representation of the dipolar gas confined longitudinally by an optical lattice and transversely by a harmonic trap. (b) Modified dipole-dipole interaction vs lattice site for different $\beta_{\text{eff}}$ (see main text), notice the log scale in the vertical axis.}
\label{Fig-setup}
\end{figure}


This manuscript is organized as follows. In Sec.~\ref{Model}, we introduce the lattice model considered, and show how the transversal confinement modifies the inter-site interaction decay. In Sec.~\ref{PI}, we study the effects of the modified interaction on the phase diagram of repulsive hard-core bosons, whereas  Sec.~\ref{NI} focuses on the attractive case.  Finally, in Sec. \ref{Conclusions}, we summarize our conclusions.

\section{Modified interaction decay}
\label{Model}
We consider dipolar hard-core bosons of mass $m$ confined longitudinally by a 1D optical lattice, $U_{0}\sin^2(\pi z/a)$ with $a$ the lattice constant, and transversally by an isotropic harmonic potential $\frac{1}{2}m\omega_{\perp}^{2}(x^2+y^2)$, $\omega_{\perp}$ being the trap frequency~(see Fig. \ref{Fig-setup}(a)). The dipole moments are assumed to be oriented by an external field on the $xz$ plane forming an angle $\alpha$ with the lattice axis $z$. Within the tight-binding approximation, the system is well described by the extended Hubbard model~(EHM):
\begin{equation}
\hat{H} \!=\! \sum_{i}
\left [
-t \left (\hat{b}^{\dagger}_{i+1}\hat{b}_{i}
+\mathrm{H.c.}\right )
\!+\!\sum_{j>0}V_{j}\hat{n}_{i}\hat{n}_{i+j}  -\mu \hat{n}_{i} \right ],
\label{Eq1}
\end{equation}
where $t$ is the tunneling amplitude between nearest neighbors, $\hat{b}_{i}^{\dagger}$ ($\hat{b}_{i}$) is the creation (annihilation) operator at site $i$, $\hat{n}_{i}=\hat{b}_{i}^{\dagger}\hat{b}_{i}$ is the corresponding particle number operator, and $\mu$ is the chemical potential. The hard-core constraint means that no double occupancy is allowed, i.e. $(\hat{b}_{i}^{\dagger})^{2}=0$. This restriction can be achieved by means of strong-enough on-site interactions (which may demand the use of Feshbach resonances). The inter-site interaction between dipoles separated by $j$ sites is given by:
\begin{equation}
V_{j} = \int d^{3}r \int d^{3}r' \ V(\vec{r}-\vec{r}')|\phi(\vec{r})|^2|\phi(\vec{r}-ja\vec{e}_{z})|^2,
\label{Eq2}
\end{equation}
with $V(\vec{r}) = \frac{C_{dd}}{4\pi r^{3}}\left(1-3\frac{(x\sin\alpha+z\cos\alpha)^{2}}{r^{2}}\right)$ being the dipole-dipole interaction. The coupling constant $C_{dd}$ is $\mu_{0}\mu^{2}$ for particles having a permanent magnetic dipole moment $\mu$ ($\mu_{0}$ is the permeability of vacuum) and $d^{2}/\varepsilon_{0}$ for particles having a permanent electric dipole moment $d$ ($\varepsilon_{0}$ is the vacuum dielectric constant). 

The on-site wave function can be written as $\phi(\vec{r})=\psi_{0}(x,y)W(z)$, where $W(z)$ is the Wannier function associated with the lowest-energy band, and $\psi_{0}(x,y) = \frac{e^{-(x^2+y^{2})/2\ell_{\perp}}}{\sqrt{\pi}\ell_{\perp}}$ is the ground-state wave function of the transversal confinement, with $\ell_{\perp} = \sqrt{\hbar/m\omega_{\perp}}$ the harmonic oscillator length. 
For deep-enough lattices, $s\gtrsim 5$, where $s=U_{0}/E_{R}$ is  the lattice depth, and $E_{R}=\frac{\pi^{2}\hbar^{2}}{2ma^{2}}$ the recoil energy, the actual Wannier function can be well approximated~(for the purpose of the calculation of the inter-site interactions) by a Gaussian~\cite{Bloch2008}, $W(z)\simeq \frac{e^{-z^{2}/2\ell}}{\sqrt{\sqrt{\pi}\ell}}$, with $\ell = a/(\pi s^{1/4})$ the effective on-site harmonic oscillator length.
After some straightforward algebra, one can show that for $\ell_{\perp} > \ell$~\cite{Korbmacher2023}:
\begin{equation}
\frac{V_{j}}{E_{R}} = \frac{3B^{3/2}}{2\pi^{2}}(3\cos^{2}\alpha-1)\left(\frac{a_{dd}}{a}\right)f(\sqrt{B}j),
\label{Eq3}
\end{equation}
where $a_{dd} = mC_{dd}/(12\pi\hbar^{2})$ is the dipolar length, $B=\frac{\pi^{2}}{2}\frac{\chi}{1-\frac{\chi}{2\sqrt{2}}}$, $\chi=\hbar\omega_\perp/E_R$, and
\begin{equation}
f(\xi) = 2\xi -\sqrt{2\pi}(1+\xi^2)e^{\xi^2/2}\text{erfc}(\xi/\sqrt{2}).
\label{Eq4}
\end{equation}
By using $V=V_{1}$, the interaction potential in Eq.~\eqref{Eq3} can be written as $V_{j}=VG_{j}(B)$, with $G_{j}(B) = f(\sqrt{B}j)/f(\sqrt{B})$. The resulting interaction decay has hence a universal dependence on the parameter $B$, which is a function of the confinement parameters only. 

Although for sufficiently large distances the potential in Eq.~\eqref{Eq3} recovers the standard dipolar tail, i.e. $V_{j\rightarrow\infty}/V \rightarrow 1/j^{3}$, the modification of the interaction to the neighboring sites may be very significant~(see Fig.~\ref{Fig-setup}(b)). The modification of the form of the tail $V_j/V$ is uniquely defined once the ratio $V_2/V$ is fixed. Below we parameterize such a ratio~(for any value of $l_\perp/l$) by means of the effective exponent $\beta_{\text{eff}}$, defined as $V_2/V=1/2^{\beta_{\text{eff}}}$~\cite{Korbmacher2023}. 
The ground-state properties are 
then fully determined by the 
value of $V/t$~(which sets the interaction strength) and $\beta_{\text{eff}}$~(which fixes the form of the decay $V_j/V$). 
By changing the transversal confinement,   $\beta_{\text{eff}}$ can be tuned within the interval $0<\beta_{\text{eff}}<3.2$~\cite{Korbmacher2023}. When $\ell_{\perp} < \ell$, $\beta_{eff}>3$, whereas the opposite is true if $\ell_{\perp} > \ell$. 
When $\beta_{\text{eff}}=3$, the decay corresponds exactly to the standard $V_j/V=1/j^3$, and hence we employ 
interchangeably $1/j^3$ or $\beta_{\text{eff}}=3$ 
to denote the standard decay.
For $\ell_{\perp} \ll \ell$, 
the case discussed in Ref.~\cite{Wall_2013}, $\beta_{eff}$ is only slightly larger than $3$, and hence the modification to the $1/j^3$ dependence is small. As a result the corrections to the ground-state properties are minor~\cite{Wall_2013}. In contrast, the deviation from the standard decay may be large when $\ell_{\perp} > \ell$, leading to a very significant modification of the ground-state properties, as discussed below.


\section{Repulsive polar lattice gas}
\label{PI}

We focus first on the ground-state properties for the case of repulsive inter-site interactions, $V>0$. In absence of dipolar interaction, the standard Hubbard model with hard-core bosons may present only two phases, either a superfluid~(SF) phase or a band insulator, with filling factor $\bar n=N/L=1$, with $N$ the number of bosons and $L$ the number of sites. Note that the latter is equivalent to the vacuum~($\bar n=0$) due to particle/hole symmetry. In the presence of inter-site dipolar interactions, and depending on the dipole strength $V/t$ and the chemical potential $\mu$ the system may present different insulating phases with commensurate fractional fillings~(devil's staircase)~\cite{Burnell_2009}. 
Particularly relevant are the 
half-filled density-wave~(2DW)~($\bar n=1/2$), which for $t=0$ acquires the form $|\cdots\,\fullcirc\,\emptycirc\,\fullcirc\,\emptycirc\,\cdots\rangle$, and the one-third-filled~(3DW)~($\bar n =1/3$) 
$|\cdots\,\fullcirc\,\emptycirc\,\emptycirc\,\fullcirc\,\emptycirc\,\emptycirc\,\cdots\rangle$~(or equivalently the phase with $\bar n=2/3$). Other fractional fillings are possible but they require significantly larger $V/t$ ratios. 

We are interested in how the modified interaction decay alters the boundaries of the insulating phases. We employ density-matrix renormalization group~(DMRG) techniques to obtain the ground-state of a system of $L=120$ sites, assuming periodic boundary conditions. Superfluid~(insulating) phases are characterized by a polynomial~(exponential) decay of the single-particle
correlation, 
$C_\text{SF}(i,j)=\langle \hat{a}_i^\dagger\hat{a}_{i+j}\rangle$. To distinguish amongst the different insulating phases, we evaluate the structure factor 
$M(k)=\frac{1}{L}\sum_{j=1}^{L-1}e^{-ikj}\langle\hat{n}_i\hat{n}_{i+j}\rangle$, 
where $k\in[-\pi, \pi]$ is the quasi-momentum. For an insulating phase with filling $\bar{n}=\frac{1}{m}$, $M(k)$ presents peaks 
at $k=\pm\frac{2\pi}{m}$. Due to particle-hole symmetry, the same is true for the phase with $\bar{n}=\frac{m-1}{m}$. Note that particle-hole symmetry results in a mirror symmetry of the phase diagram on the 
$(t/V,\mu/V)$ plane around the chemical potential $(\mu/V)_0(B)=\sum_{j>0}G_j(B)$. 
In order to compare properly the results for different $B$ values, we introduce the re-scaled chemical potential, $\tilde{\mu}=\mu/(2\sum_{j>0}G_j(B))$, such that the phase diagram presents mirror symmetry around $\tilde{\mu}/V=\frac{1}{2}$. 
With this shift, the borders of the band insulator, $\bar{n}=1$, and the 
vacuum, $n=0$, are given by $\tilde{\mu}/V=0,1$. 



\begin{figure}[t]
\centering
\includegraphics[width=0.8\columnwidth]{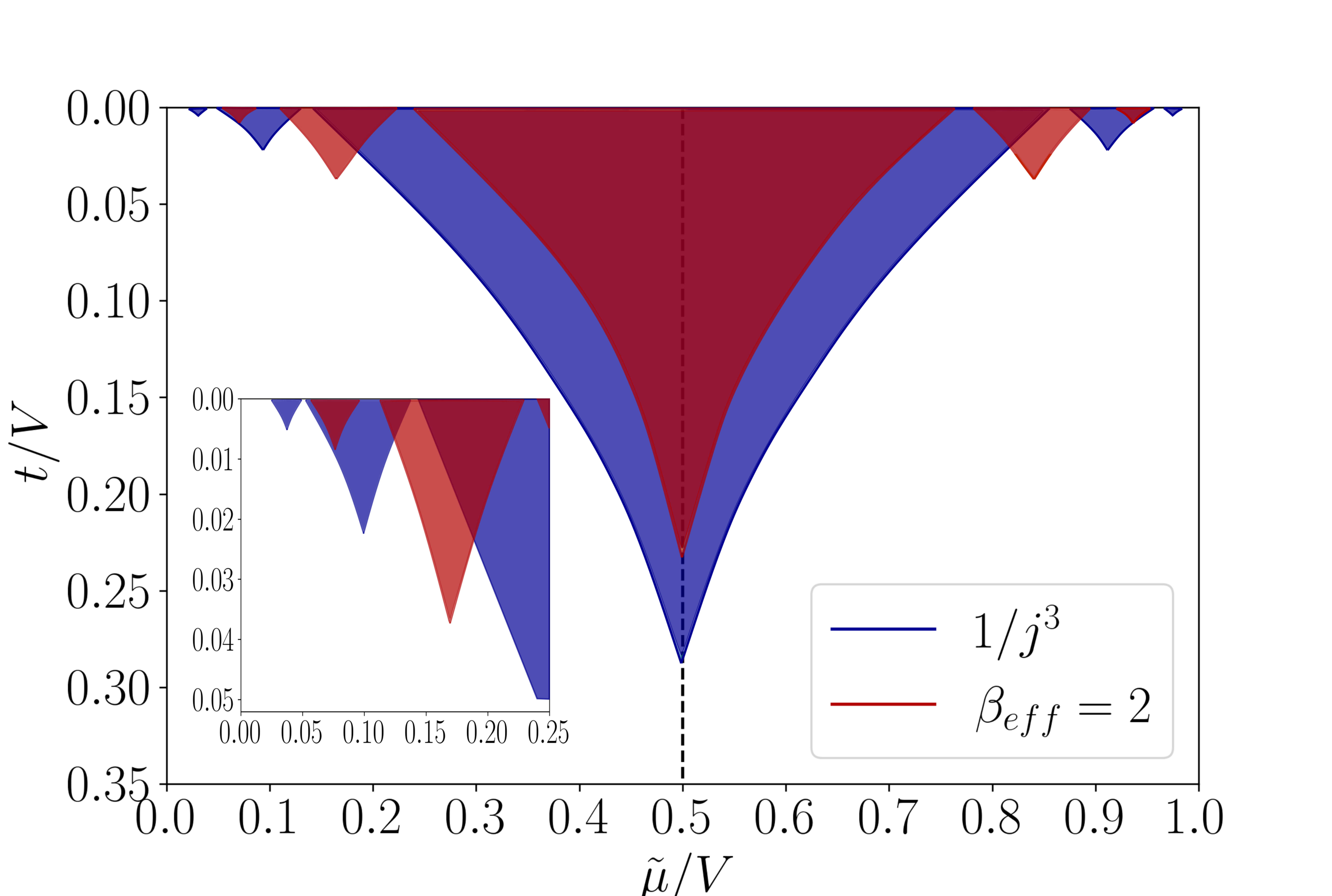}
\includegraphics[width=0.8\columnwidth]{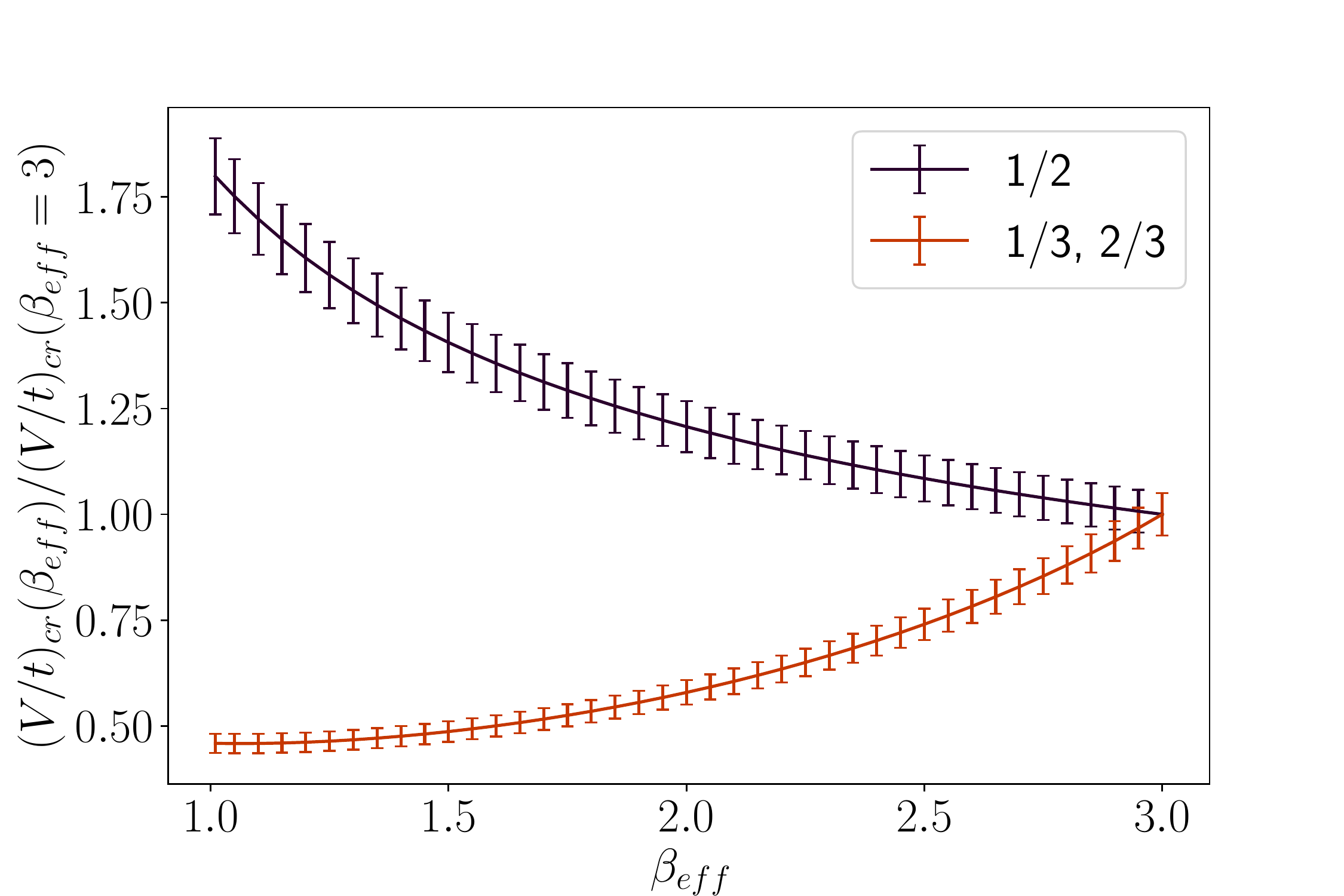}
\caption{(top) Phase diagram in the $(t/V,\tilde{\mu}/V)$  plane for standard $1/j^3$ decay~(blue lobes) and the modified dipolar interaction with $\beta_{\text{eff}}=2$ (red lobes). (bottom) Critical $V/t$ for the 2DW~(blue) and the 
3DW~(red) for different values of $\beta_\text{eff}$, normalized to the value expected for the $1/j^3$ decay. Error bars indicate the exact value of $(V/t)_{cr}$ and are set by the fidelity of numerical input parameters.}
\label{Fig-PhaseDiagram}
\end{figure}


In Fig.~\ref{Fig-PhaseDiagram}~(top), we depict the phase diagram for a 
$1/r^{3}$ decay~(blue), and for the modified dipolar interaction with $\beta_{\text{eff}}=2$~(red). 
Due to the enhanced role of the next-to-NN interaction $V_2$ compared to the usual $1/j^3$ decay,
the central lobe, which corresponds to the 2DW phase, is significantly smaller for $\beta_\text{eff}=2$. This must be compared to the results of Ref.~\cite{Wall_2013}, which focused on the case $\ell_{\perp} \ll \ell$~(and hence $\beta_{\mathrm{eff}}\gtrsim 3$), for which the modification of the 2DW lobe compared to that expected for the $1/j^3$ dependence is very small. 

The relative deviation from the results considering
the standard $1/j^3$ dependence is even more relevant 
for insulating phases at lower fillings. In Fig.~\ref{Fig-PhaseDiagram}~(top) we observe as well the lobes with filling $\bar n=1/3$ and $\bar n=1/4$, and the particle/hole symmetric ones for $\bar n=2/3$ and $\bar n=3/4$. Note that these lobes, which are depicted in detail in the inset, are significantly modified, 
again due to the enhanced value of $V_2$ and $V_3$ compared to the standard $1/j^3$ decay.
In particular, for $\beta_{\mathrm{eff}}=2$, the critical $V/t$ for the observation of the 3DW is strongly reduced from a critical $V/t \simeq 47$ to $V/t\simeq 27$. Figure~\ref{Fig-PhaseDiagram}~(bottom) shows the critical $(V/t)_{cr}$ for 2DW and 3DW, as a function of $\beta_\text{eff}$, normalized to the value expected for a $1/j^3$ decay. Note that for 2DW~(3DW) $(V/t)_{cr}$ increases~(decreases) by 
approximately a factor of $2$ when reducing $\beta_{\mathrm{eff}}$ down to $1$.



\begin{figure}[t]
\centering
\includegraphics[width=0.8\columnwidth]{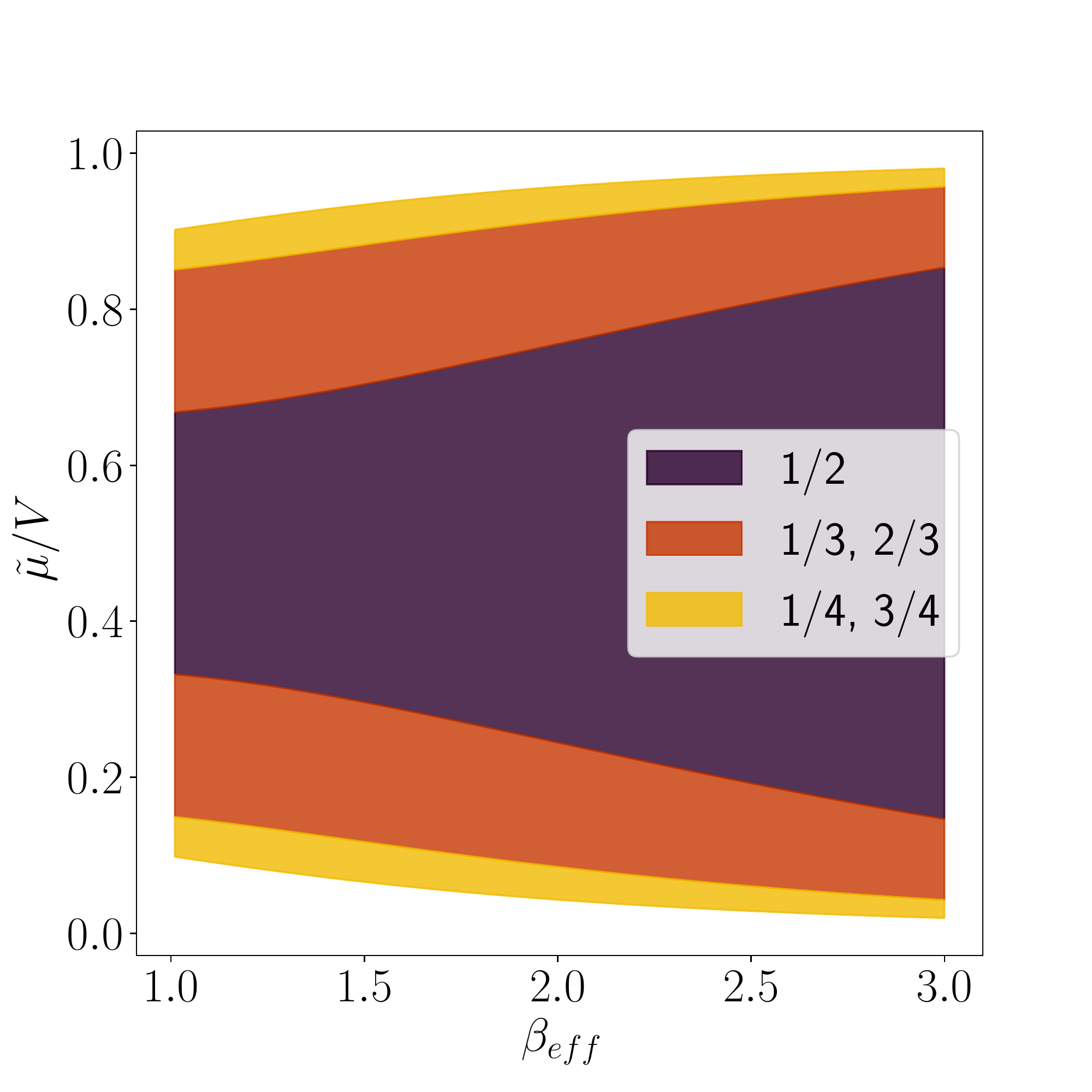}
\includegraphics[width=0.8\columnwidth]{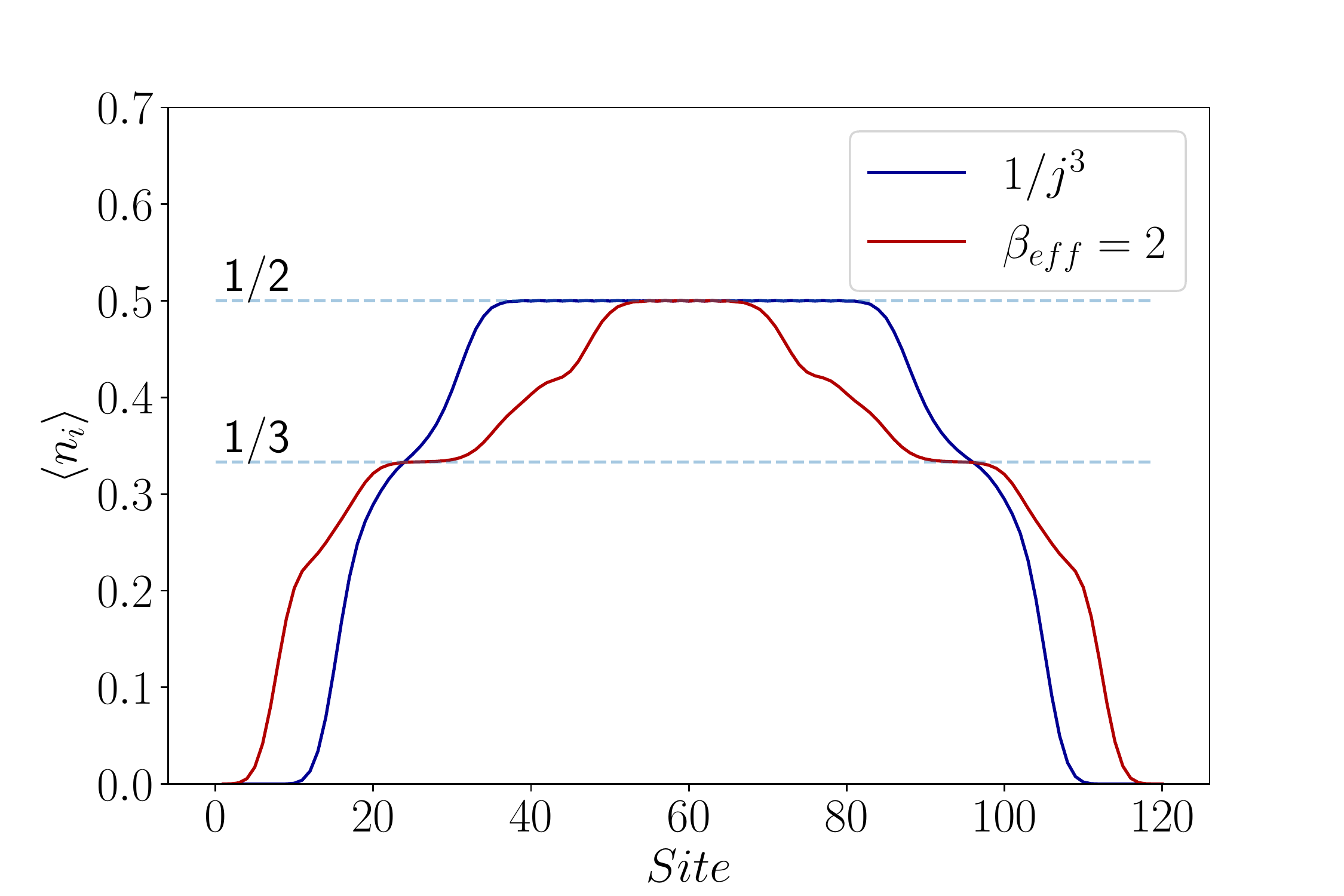}
\caption{(top) Boundaries of the lobes with $n=1/2$, $1/3$ and $1/4$ for $t=0$ as a function of the shifted chemical potential $\tilde\mu/V$ and $\beta_\text{eff}$. The lobes are evaluated analytically considering a cut-off of the interactions at $4$ neighbors. (bottom) 
Spatial density distribution $\langle n_i\rangle$ for $N=39$ bosons, $\Omega/t=0.03$, and for $1/j^3$ decay~(blue) and for $\beta_\text{eff}=2$~(red) obtained from DMRG calculations.
}
\label{Fig-t0}
\end{figure}


The lobes are also changed in their dependence with chemical potential. In addition to the above-mentioned $(\mu/V)_0(B)$ shift, the width~(in chemical potential) of the 2DW~(3DW) lobe significantly shrinks~(widens) with decreasing $\beta_{\mathrm{eff}}$. This is illustrated in Fig.~\ref{Fig-t0}~(top), where 
we consider $t=0$, for which the 
phase boundaries may be evaluated analytically. 

The modification of the insulating lobes has significant consequences 
for the spatial particle distribution in the 
the presence of an overall harmonic confinement.
The confinement results in an additional term $\Omega\sum_i(i-L/2)^2\hat{n}_i$ in 
Eq.~\eqref{Eq1}. For a sufficiently weak confinement, local-density approximation 
arguments apply, and the density profile presents the expected wedding-cake profile. Figure~\ref{Fig-t0}~(bottom) shows the local mean occupation $\langle n_i\rangle$ obtained from DMRG calculations. In the plateaus, which characterize 
the DW phases, we average over neighboring sites to flatten the DW oscillations in $\langle n_i\rangle$. In agreement with the phase diagram, the central 2DW plateau shrinks while the 3DW plateau widens when decreasing $\beta_{\text{eff}}$.



\section{Attractive polar lattice gas}
\label{NI}

We analyze at this point the case of  $V<0$, focusing first on the simplest case of just two bosons, and then discussing the formation of self-bound lattice droplets.

\subsection{Dimers}

\begin{figure}[t]
\centering
\includegraphics[width=\columnwidth]{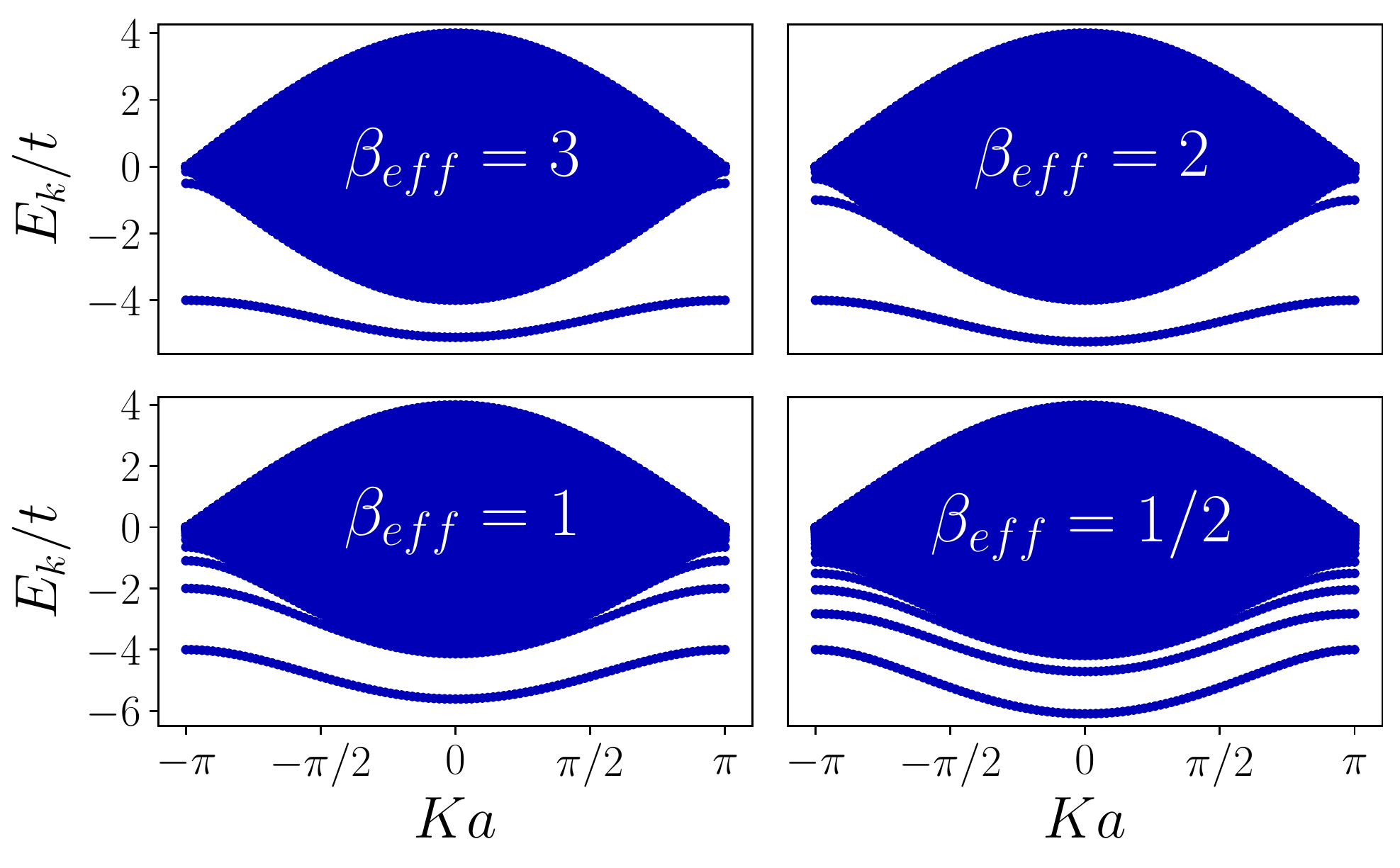}
\caption{Energy spectrum $E_{K}$ as a function of the center-of-mass quasi-momentum $K$ for several effective powers of the dipole-dipole interaction. In all panels the interaction strength is $V/t=-4$.}
\label{Figure6}
\end{figure}

The wave function characterizing a state of two bosons can be separated as $\Psi(R, r) = e^{iKR}\Phi_{K}(r)$, where $R=(i_{1}+i_{2})/2$ is the center-of-mass, $r=i_{1}-i_{2}$ is the relative coordinate, and $i_{j=1,2}$ is the lattice site in which particle $j$ is. The wave function $\Phi_{K}(r)$ depends on the center-of-mass quasi-momentum $K\in[-\pi, \pi]$, and satisfies the Schr\"odinger equation $\hat{H}_{K}|\Phi_{K}\rangle = E_{K}\Phi_{K}$, with 
\begin{eqnarray}
\hat{H}_{K} &=& -2t\cos\left (\frac{Ka}{2} \right )\sum_{r\geq 1} (|K, r+1\rangle\langle K, r| + \mathrm{H.c.}) \nonumber \\
&+& V\sum_{r\geq 1}G_{j}(B)|K,r\rangle\langle K, r|,
\label{Eq7}
\end{eqnarray}
where $|K, r\rangle$ stands for the state with center-of-mass quasi-momentum $K$, and inter-particle separation $r$. Diagonalizing $\hat{H}_{K}$ for different $K$ 
in the Brillouin zone provides the energy spectrum,  
depicted in Fig.~\ref{Figure6} for $V/t=-4$ and different values of $\beta_{\text{eff}}$. Compared to the case $1/r^3$ for the same $V/t$, the modification of the dipolar tail results in additional bound eigenstates. 

The two-body ground-state is given for all $\beta_{\mathrm{eff}}$ values by a bound pair  with $K=0$, and a spatial distribution peaked at nearest neighbors. However, the binding becomes stronger, i.e. pairing (and in general the formation of clusters, as discussed below) demands a smaller $|V|/t$ when $\beta_{\mathrm{eff}}$ decreases. 
Furthermore, dimer mobility may be strongly modified by tuning the transversal confinement. 
Note that the curvature of the lowest branch at $K=0$, associated with the effective mass of the ground-state dimer, is significantly modified as a function of $\beta_{\mathrm{eff}}$. 
This is best illustrated with the case of strong $V/t$, for which the lowest bound-state branch corresponds to bound nearest-neighbor dimers, which 
move via second-order hopping with amplitude 
\begin{equation}
t_{D}=\frac{1} {1-2^{-\beta_{\mathrm{eff}}}} \frac{t^2}{V}.
\end{equation}
This should be compared to the corresponding value $t_{D} = 8t^{2}/7V$ for the case of $1/r^3$ decay. Note that e.g. for $\beta_{\text{eff}}=1$, $t_{D} = 2 t^{2}/V$, and hence the dimer dynamics is approximately twice faster.


\subsection{Self-bound lattice droplets}

The formation of bound dimers in the two-body problem extends in the many-body case to the formation of self-bound lattice droplets formed by potentially many particles. These resemble those recently discussed in binary mixtures~\cite{PhysRevResearch.2.022008, PhysRevLett.126.023001} and cavities~\cite{Karpov2019, Karpov2022}. Dipolar self-bound lattice droplets have been discussed in the context of out-of-equilibrium polar lattice gases after a quench of the confinement potential~\cite{Li2020}. Very recently, ground-state self-bound dipolar lattice droplets were studied in Ref.~\cite{Morera2023}.
Although self-bound lattice droplets present some interesting similarities to quantum droplets in binary and dipolar Bose-Einstein condensates~\cite{Boettcher2021}, they differ from them 
in the physical mechanism as well as in the fact that lattice droplets are self-pinned, i.e. they remain for any practical purposes immobile, due to their large effective mass.

In Ref.~\cite{Morera2023}, it was shown that droplets 
can be either a self-bound Mott insulator~(with saturated unit filling) or in a liquefied state~(self-bound but with a filling lower than unity). It was argued that liquefaction arises due to the interplay between inter-site dipolar attraction and the super-exchange processes originating from short-range repulsion in soft-core Bose systems. 
In the following, we show that self-bound droplets in hard-core gases (where super-exchange is absent) are generally in either a saturated or liquid regime, and that the boundaries between the saturated, liquid, and unbound~(gas) phases are 
strongly dependent on $\beta_{\mathrm{eff}}$.



\begin{figure}[t!]
\centering
\includegraphics[width=1.0\columnwidth]{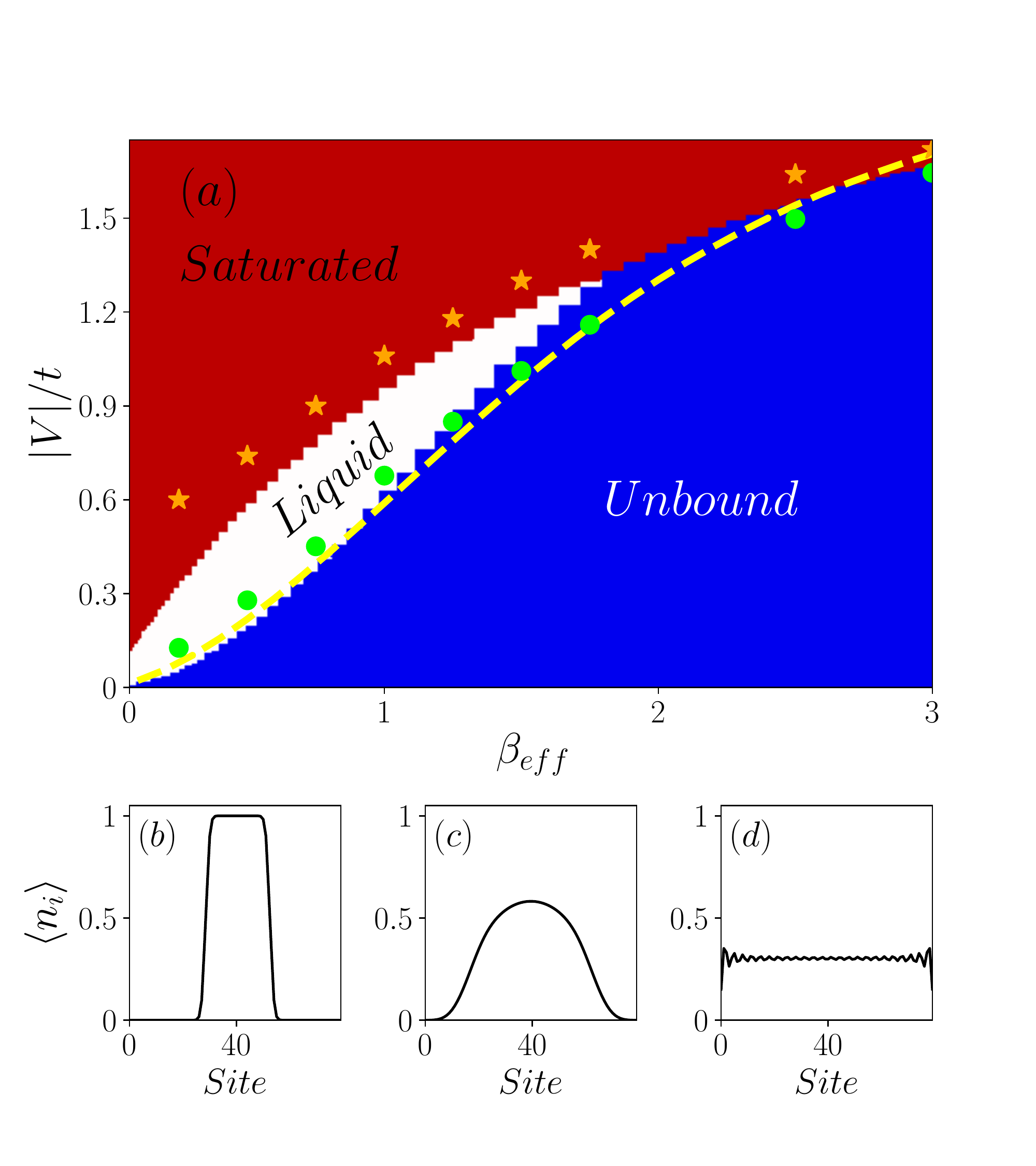}
\caption{(a) Phase diagram for impenetrable dipolar bosons as a function of the absolute value of the interaction strength $|V|/t$ and $\beta_{\text{eff}}$. We found a saturated droplet regime (b), a droplet superfluid (c), and a unbound phase (gas) (d). The Tonks-Girardeau analysis corresponds to the color scheme shown, saturated
droplet regime (red), droplet superfluid (blue), and gaseous phase (white). Orange and green markers are associated with the boundary lines of the saturated droplet and liquid regimes from DMRG calculations, respectively. The dashed yellow line indicates the threshold for a two-body bound state (dimer). Lower panels display characteristic density profiles of the different phases.}
\label{Figure8}
\end{figure}


In order to analyze self-bound lattice droplets, we develop a 
variational  approach similar to that used in Ref.~\cite{Morera2023}. We assume that the impenetrable lattice gas is well described by a Tonks-Girardeau $|\Psi_{TG}\rangle=\Pi_{k<k_{F}}\hat{b}_{k}^{\dagger}|0\rangle$ ansatz with the density $n=k_{F}/\pi$ as a variational parameter, $k_{F}$ being the Fermi momentum of the fermionized bosons. Although this ansatz is only exact for  non-interacting one-dimensional hard-core bosons, it can perturbatively capture the main features of dipolar hard-core gases~\cite{Morera2023}. Evaluating  $\langle\Psi_{TG}|\hat{H}|\Psi_{TG}\rangle$ yields the energy per particle
\begin{equation}
\frac{E[n]}{tN} = -\frac{2\sin n\pi}{n\pi}+\frac{V}{tn}\sum_{r>0}G_{r}\left[1-\frac{\sin^{2}n\pi r}{\pi^{2}r^{2}}\right].
\end{equation}
The first and second terms are associated with the kinetic energy and the modified dipolar interaction, respectively. Following Ref.~\cite{Morera2023}, we classify the quantum phases of the dipolar system according to the value of the density $n_{c}$ at which the energy per particle is minimal. In the unbound~(gas) phase, the dipoles spread uniformly over all available sites. As a result, the gas phase is characterized by a vanishing density $n_{c}=0$. In contrast, in a liquid phase, where the droplets are self-bound and localized at zero pressure \cite{PhysRevResearch.2.022008, PhysRevLett.126.023001}, the energy per particle takes its minimal value at a finite density $0<n_{c}<1$. Furthermore, the energy per particle at $n_{c}$ is smaller than the bottom of the scattering band, i.e $E[n_{c}]/N<-2t$. Lastly, we define the saturated droplet regime as that in which the energy per particle becomes minimal at $n_{c}=1$. In such a phase, the droplet is incompressible and hole propagation within the droplet is inhibited due to the high energy cost of breaking a dipole bond \cite{Li2020}. In Fig. \ref{Figure8}, we show, in a color scheme, the resulting phase diagram of impenetrable dipolar bosons as a function of the interaction strength $|V|/t$ and $\beta_{\text{eff}}$. The saturated droplet regime is indicated in red, whereas the liquid and gaseous phases in white and blue, respectively. In stark contrast to the bare dipolar potential $1/r^{3}$~\cite{Morera2023}, the modified interaction gives rise to a wide liquefied region without removing the hard-core constraint.

In addition to the Tonks-Girardeau analysis, we calculate the ground state of the dipolar system for different values of $|V|/t$ using DMRG simulations with $N=24$ bosons in $L=80$ sites. In our DMRG simulations, we define the gas-to-liquid transition at the interaction strength $|V_{c}|/t$ in which the energy per particle is equal to the bottom of the scattering band $E/N = -2t$. Meanwhile, we define the saturated droplet regime as that in which the central density of the ground state reaches unity, i.e. the density distribution acquires a flat-top profile. Note that the wings of the droplet are not necessarily saturated, although when $|V|/t$ increases eventually the whole droplet enters the unit filling regime. Green circles and orange stars in Fig.~\ref{Figure8} correspond to our DMRG results for the gas-to-liquid and liquid-to-saturated boundaries, respectively. A good agreement is found between the Tonks-Girardeau analysis and the DMRG results. Note as well, that the gas-to-liquid transition is well-estimated by the threshold of dimer bound-state formation (dashed line in Fig.~\ref{Figure8}), obtained from the two-body calculations discussed above.


\section{Conclusions}
\label{Conclusions}
A sufficiently loose transversal confinement  results in a significant 
modification of the inter-site interaction between dipoles in a one-dimensional optical lattice, which departs from the usually assumed 
$1/j^3$ dependence. We have shown that this modification, which 
acquires a universal dependence on the confinement parameters, may significantly modify the ground-state properties of hard-core bosons. For repulsive dipoles, it leads to a marked shift of the boundaries of the insulating devil's staircase phases, that translates in a significantly modified particle distribution in the presence of an overall harmonic potential. For attractive dipoles, the modified interaction decay results in a lower critical dipolar strength for the formation of self-bound clusters, 
and to a much wider parameter region for the observation of liquefied droplets without the need of super-exchange processes. The discussed effects should play a relevant role in future lattice experiments on magnetic atoms or polar molecules.

\section*{Acknowledgments}
We acknowledge support of the Deutsche Forschungsgemeinschaft (DFG, German Research Foundation) -- Project-ID 274200144 -- SFB 1227 DQ-mat within the project A04, and under Germany's Excellence Strategy -- EXC-2123 Quantum-Frontiers -- 390837967. J.Z. is supported by the National Science Centre (Poland) under project   2021/43/I/ST3/01142 founded in the OPUS call within the WEAVE programme.

\bibliography{Dipolar_Version1}
\end{document}